\documentclass[12pt]{article}
\usepackage{graphicx}
\def\beq{\begin{equation}}
\def\eeq{\end{equation}}
\def\bea{\begin{eqnarray}}
\def\eea{\end{eqnarray}}
\def\nn{\nonumber}

\def\roughly#1{\mathrel{\raise.3ex\hbox
{$#1$\kern-.75em\lower1ex\hbox{$\sim$}}}}
\def\lsim{\roughly<}

\def\sla#1{\raise.15ex\hbox{$/$}\kern-.57em #1}

\def\ket#1{\left| #1\right\rangle}
\def\ks{K_S}
\def\kbar{{\bar K}^0}
\def\bd{B_d^0}
\def\bs{B_s^0}

\def\btod{{\bar b} \to {\bar d}}
\def\btos{{\bar b} \to {\bar s}}
\def\order{\lower 1.8ex \hbox{\LARGE\~{}}}

\def\bsbar{{\bar B}_s^0}

\makeatletter
\newcommand{\contraction}[5][1ex]{%
 \mathchoice
  {\contraction@\displaystyle{#2}{#3}{#4}{#5}{#1}}%
  {\contraction@\textstyle{#2}{#3}{#4}{#5}{#1}}%
  {\contraction@\scriptstyle{#2}{#3}{#4}{#5}{#1}}%
  {\contraction@\scriptscriptstyle{#2}{#3}{#4}{#5}{#1}}}%
\newcommand{\contraction@}[6]{%
 \setbox0=\hbox{$#1#2$}%
 \setbox2=\hbox{$#1#3$}%
 \setbox4=\hbox{$#1#4$}%
 \setbox6=\hbox{$#1#5$}%
 \dimen0=\wd2%
 \advance\dimen0 by \wd6%
 \divide\dimen0 by 2%
 \advance\dimen0 by \wd4%
 \vbox{%
  \hbox to 0pt{%
   \kern \wd0%
   \kern 0.4\wd2
   \contraction@@{\dimen0}{#6}%
   \hss}%
  \vskip 0.8ex
  \vskip\ht2}}

\newcommand{\contraction@@}[3][0.06em]{%
 \hbox{%
  \vrule width #1 height 0pt depth #3%
  \vrule width #2 height 0pt depth #1%
  \vrule width #1 height 0pt depth #3%
  \relax}}
\makeatother
\begin{document}

\begin{flushright}
UdeM-GPP-TH-11-196 \\
\end{flushright}

\begin{center}
\bigskip
{\Large \bf \boldmath SU(3) Breaking in Charmless $B$ Decays} \\
\bigskip
\bigskip
{\large 
Maxime Imbeault $^{a,}$\footnote{imbeault.maxime@gmail.com}
and David London $^{b,}$\footnote{london@lps.umontreal.ca}
}
\end{center}

\begin{flushleft}
~~~~~~~~~~~$a$: {\it D\'epartement de physique, C\'egep de Baie-Comeau,}\\
~~~~~~~~~~~~~~~{\it 537 boulevard Blanche, Baie-Comeau, QC, Canada G5C 2B2}\\
~~~~~~~~~~~$b$: {\it Physique des Particules, Universit\'e
de Montr\'eal,}\\
~~~~~~~~~~~~~~~{\it C.P. 6128, succ. centre-ville, Montr\'eal, QC,
Canada H3C 3J7}\\
\end{flushleft}

\begin{center}
\bigskip (\today)
\vskip0.5cm {\Large Abstract\\} \vskip3truemm
\parbox[t]{\textwidth}{There are many charmless $B$ decay pairs whose
  amplitudes are related by U spin ($d\leftrightarrow s$) or flavor
  SU(3). The theoretical uncertainty in any analysis involving such
  pairs must take into account U-spin/SU(3) breaking. In the past,
  such considerations generally used theoretical input, but we show
  that this can be experimentally measured. We present lists of two-
  and three-body decay pairs from which the size of the breaking can
  be obtained. We detail the values of U-spin/SU(3) breaking given by
  the present experimental data. One pair -- $\bd \to \pi^+\pi^-$ and
  $\bd \to\pi^-K^+$ -- exhibits large nonfactorizable breaking. We
  present other signals of SU(3) breaking in two- and three-body
  decays, and discuss further tests for nonfactorizable
  effects. Finally, we also point out that the pure-penguin decay $\bs
  \to \kbar \kbar K^0$ is intriguing because it can be used to cleanly
  probe the $\bs$-$\bsbar$ mixing phase.  }
\end{center}

\thispagestyle{empty}
\newpage
\setcounter{page}{1}
\baselineskip=14pt

\section{Introduction}

In the standard model (SM), CP violation is due to the presence of a
nonzero complex phase in the Cabibbo-Kobayashi-Maskawa (CKM) quark
mixing matrix $V$.  This phase information is elegantly displayed in
the unitarity triangle, in which the CP-violating interior angles are
$\alpha$, $\beta$ and $\gamma$ \cite{pdg}. By measuring these CP
phases in many different ways, one can test the SM.

Much theoretical work has gone into elucidating clean methods for
obtaining $\alpha$, $\beta$ and $\gamma$ from $B$ decays. In 1999, it
was pointed out that, apart from CKM matrix elements, the amplitudes
for the decays $\bd\to\pi^+\pi^-$ and $\bs\to K^+ K^-$ are equal under
U-spin symmetry ($d\leftrightarrow s$) \cite{Fleischer99}. With one
additional piece of information, the phase $\gamma$ can be
obtained. Subsequently, all $B$ decay pairs that are related by U spin
were tabulated \cite{GroUspin}, and another method for extracting
weak-phase information using a different U-spin pair ($\bs \to \pi^+
K^-$ and $\bd \to \pi^- K^+$) was proposed \cite{GRBsKpi}.

In order to determine the theoretical uncertainty of a particular
method, it is necessary to address the issue of U-spin breaking. In
general, theoretical input is used. However, one of the purposes of
the present paper is to note that, in fact, this can be experimentally
measured. The point is that, under U-spin symmetry, four of the
experimental observables -- the branching ratios and direct CP
asymmetries of the two decays -- are related, i.e.\ they are not
independent. Thus, the experimental values of these observables, and
the extent to which the relation among them is not satisfied, gives a
measure of U-spin breaking. Note: this is not a completely new
result. The relation among the four observables already appears in a
number of papers. However, in general it is used as a theoretical
constraint, rather than an experimental result.

In addition, one can go further. If one neglects annihilation- and
exchange-type diagrams (which are expected to be small) in the $B$
decay amplitudes, there are other pairs of amplitudes which are equal,
apart from CKM matrix elements \cite{GHLR}. In this case, it is not U
spin that is assumed, but rather full flavor SU(3)
symmetry\footnote{Note: because isospin is a good symmetry, in
  practice there is little difference between U spin and SU(3).}. Here
there are many more pairs whose amplitudes are related. And because
the relation among the four observables holds in the SU(3) limit, it
is possible to measure SU(3)-breaking effects using any of these decay
pairs.

In fact, there are a number of two-body $B$ decay pairs for which this
information is presently available. Furthermore, in such decays, the
factorizable contribution to the breaking is often under good
theoretical control. If this is taken into account, the measurement of
U-spin/SU(3) breaking then tells us the size of nonfactorizable
effects. In most cases, the data shows that such effects are small.
However, as we show below, there is one decay pair -- $\bd \to
\pi^+\pi^-$ and $\bd \to\pi^-K^+$ -- which exhibits large
nonfactorizable breaking.  Although this is just one data point, so
that no strong conclusions should be drawn, it does raise questions
about analyses which neglect nonfactorizable U-spin/SU(3) breaking.

We begin in Sec.~2 with a discussion of U spin and U-spin breaking as
it applies to a pair of charmless $\btod$ and $\btos$ decays . We show
how the measurement of the branching ratios and direct CP asymmetries
of these two decays allows one to experimentally measure the breaking.
In Sec.~3, we turn to an examination of two-body $B$ decays. We
present lists of 5 U-spin pairs and 11 additional SU(3) pairs whose
U-spin/SU(3) breaking can be measured using this method. We show the
latest data for five of these pairs. For two of these, the
measurements are reasonably accurate, and one pair shows signs of
significant nonfactorizable U-spin/SU(3) breaking. Finally, we discuss
several pairs of decays whose amplitudes are equal, including CKM
factors, within SU(3). A measure of SU(3) breaking is given by
comparing the branching ratios of the two decays, as well as the
direct CP asymmetries.

We discuss three-body decays in Sec.~4.  There are 7 decay pairs whose
amplitudes are related by U spin -- the amount of breaking can be
measured experimentally using the above method. In passing, we note
that the pure-penguin decay $\bs \to \kbar \kbar K^0$ is
interesting. Given that the final state is a CP eigenstate, the
measurement of the indirect CP asymmetry in this decay cleanly probes
the $\bs$-$\bsbar$ mixing phase, and might be easier experimentally
than what is done at the moment.  We also present the list of an
additional 24 decay pairs whose amplitudes are related by SU(3). In
this case, all final-state particles are identical, and so
permutations of these particles must be considered. We show that, in
(almost) all cases, the amplitudes are equal only for the totally
symmetric final state $|S\rangle$, so that this state must be isolated
experimentally in order to measure SU(3) breaking. We also point out
the decay pairs whose amplitudes are equal, including CKM factors,
within SU(3) for $|S\rangle$. In principle, these can also give
information about SU(3) breaking. We conclude in Sec.~5.

\section{\boldmath U Spin and U-Spin Breaking}
\label{Uspinbreak}

Consider charmless $\btod$ and $\btos$ decays\footnote{Much of the
  material in this section can be found in
  Refs.~\cite{GroUspin,NSL}.}. Their amplitudes can be written as
\bea
A(\btod) &=& A_u \lambda_u^{(d)} + A_c \lambda_c^{(d)} + A_t \lambda_t^{(d)} ~, \nn\\
A(\btos) &=& A'_u \lambda_u^{(s)} + A'_c \lambda_c^{(s)} + A'_t \lambda_t^{(s)} ~,
\label{amps}
\eea
where the $A_i$ and $A'_i$ ($i=u,c,t$) each represent a linear
combination of diagrams, and $\lambda_p^{(q)}=V^*_{pb} V_{pq}$. Using
the unitarity of the CKM matrix
($\lambda_u^{(q)}+\lambda_c^{(q)}+\lambda_t^{(q)}=0$), we can reduce
the number of terms in the amplitudes from three to two. For instance,
if the $\lambda_c^{(q)}$ piece is eliminated, we have
\bea
A(\btod) &=& (A_u - A_c) \lambda_u^{(d)} + (A_t - A_c) \lambda_t^{(d)} \nn\\
&=& (A_u - A_c) \left[ |\lambda_u^{(d)}| e^{i\gamma} + \frac{(A_t -
    A_c)}{(A_u - A_c)} |\lambda_t^{(d)}| e^{-i\beta} \right] \nn\\
&=& C \left[ |\lambda_u^{(d)}| e^{i\gamma} + |\lambda_t^{(d)}| r e^{i \delta_r} e^{-i\beta} \right] ~, \nn\\
A(\btos) &=& (A'_u - A'_c) \lambda_u^{(s)} + (A'_t - A'_c)
\lambda_t^{(s)} \nn\\
&=& (A'_u - A'_c) \left[ |\lambda_u^{(s)}| e^{i\gamma} - \frac{(A'_t -
    A'_c)}{(A'_u - A'_c)} |\lambda_t^{(s)}| \right] \nn\\
&=& C' \left[ |\lambda_u^{(s)}| e^{i\gamma} - |\lambda_t^{(s)}| r' e^{i \delta'_r} \right] ~,
\label{fullamps}
\eea
where $C \equiv (A_u - A_c)$, $C' \equiv (A'_u - A'_c)$, $r e^{i
  \delta_r} \equiv (A_t - A_c)/(A_u - A_c)$ and $r' e^{i \delta'_r}
\equiv (A'_t - A'_c)/(A'_u - A'_c)$. Above we have explicitly written
the weak-phase dependence, including the minus sign from $V_{tb}^*
V_{ts}$. 

If the two amplitudes are given by a similar combination of diagrams,
then under U-spin symmetry, which exchanges $d$ and $s$ quarks, we
have $A'_i=A_i$, so that
\beq
C' = C ~~,~~~~ r' = r ~~,~~~~ \delta'_r = \delta_r ~,
\label{Uspinrels}
\eeq
and the two amplitudes are described by four unknown theoretical
parameters: $\gamma$, $|C|$, $r$, $\delta_r$ ($\beta$ has been
measured quite accurately through the indirect CP asymmetry in $\bd\to
J/\psi \ks$ \cite{pdg}, and is therefore taken to be known).

In general, there are four observables in the $\btod$ and $\btos$
processes:
\bea
B_d &=& |A(\bar b \to \bar d)|^2+|A(b \to d)|^2 ~,\nn\\
B_s &=& |A(\bar b \to \bar s)|^2+|A(b \to s)|^2 ~,\nn\\
A_d &=& \frac{|A(\bar b \to \bar d)|^2-|A(b \to d)|^2}{|A(\bar b \to \bar d)|^2+|A(b \to d)|^2}~,\nn\\
A_s &=& \frac{|A(\bar b \to \bar s)|^2-|A(b \to s)|^2}{|A(\bar b \to \bar s)|^2+|A(b \to s)|^2}~.
\label{ABdefs}
\eea
$B_d$ and $B_s$ are related to the CP-averaged $\btod$ and $\btos$
decay rates, while $A_d$ and $A_s$ are direct CP asymmetries. The
CP-conjugate amplitude $A(\bar b \to \bar q)$ is obtained from $A(b
\to q)$ by changing the signs of the weak phases.

Since there are four unknown theoretical parameters in the amplitudes
in the U-spin limit, one might imagine that these can be determined
from measurements of $B_{d,s}$ and $A_{d,s}$. However, this is not
true. It is straightforward to show that, in this limit, $X=1$, where 
\beq
X \equiv -\frac{A_s}{A_d} \, \frac{B_s}{B_d} ~.
\label{Xdef}
\eeq
Thus, there are only three independent observables.  This implies that
\beq
-\frac{|A(\bar b \to \bar s)|^2-|A(b \to s)|^2}{|A(\bar b \to \bar d)|^2-|A(b \to d)|^2} = 1 ~.
\eeq

Explicitly, we have
\beq
-\frac{|A(\bar b \to \bar s)|^2-|A(b \to s)|^2}{|A(\bar b \to \bar d)|^2-|A(b \to d)|^2} 
= \frac{|\lambda_u^{(s)}| |\lambda_t^{(s)}| \, \sin{\gamma} \, |C'|^2 r' \sin{\delta'_r}}{|\lambda_u^{(d)}| |\lambda_t^{(d)}| \, 
\sin{\alpha} \, |C|^2 r \sin{\delta_r}} ~.
\eeq
Now, the sine law associated with the unitarity triangle gives
\beq
\frac{\sin{\gamma}}{|\lambda_t^{(d)}|} = 
\frac{\sin{\alpha}}{|\lambda_c^{(d)}|} = 
\frac{\sin{\beta}}{|\lambda_u^{(d)}|} ~. 
\eeq
We therefore have
\bea
-\frac{|A(\bar b \to \bar s)|^2-|A(b \to s)|^2}{|A(\bar b \to \bar d)|^2-|A(b \to d)|^2} 
& = & \frac{|\lambda_u^{(s)}| |\lambda_t^{(s)}| \, |C'|^2 r' \sin{\delta'_r}}{|\lambda_u^{(d)}| |\lambda_c^{(d)}| \, 
|C|^2 r \sin{\delta_r}} \nn\\
&=& \frac{|V_{us}| |V_{tb}| |V_{ts}| \, |C'|^2 r' \sin{\delta'_r}}{|V_{ud}| |V_{cb}| |V_{cd}| \, |C|^2 r \sin{\delta_r}} \nn\\
&=& \frac{|C'|^2 r' \sin{\delta'_r}}{ |C|^2 r \sin{\delta_r}} ~, 
\label{Xdef2}
\eea
where $|V_{us}| |V_{tb}| |V_{ts}|/|V_{ud}| |V_{cb}| |V_{cd}| = 1$. The
above ratio equals 1 only in the U-spin limit. Thus, $(X-1)$ is a measure
of U-spin breaking. 

Until now, when this breaking was taken into account, it was only
through theoretical estimates (e.g.\ see Refs.~\cite{Fleischer99,Fleischer}).
However, in fact it can be obtained from the experimental data. This
can be combined with the theoretical calculations to look for large
nonfactorizable corrections (we will see this explicitly in
Sec.~\ref{nonfactmeasure}).  Furthermore, if the theoretical
prediction of U-spin breaking is accurate, one can use the measurement
of $(X-1)$ to search for new physics \cite{NSL}.

\section{Two-Body Decays}

\subsection{U-spin pairs}
\label{2bodyUspin}

We begin with $B\to PP$ decays ($P =$ pseudoscalar), focusing on
those $\btod$ and $\btos$ processes that are related by U spin. (It is
straightforward to extend our analysis to other two-body decays, such
as $B\to VP$ ($V =$ vector).) There are five U-spin pairs:
\begin{enumerate}

\item $\bd \to \pi^+ \pi^-$ and $\bs \to K^+ K^-$,

\item $\bs \to \pi^+ K^-$ and $\bd \to \pi^- K^+$,
	
\item $B^+ \to K^+ \bar K^0$ and $B^+ \to \pi^+ K^0$,
	
\item $\bd \to K^0 \kbar$ and $\bs \to \kbar K^0 $,
	
\item $\bd \to K^+ K^-$ and $\bs \to \pi^+ \pi^-$.

\end{enumerate}
The first (second) decay is $\btod$ ($\btos$). In all cases, the two
decays within a pair are related by U-spin reflection ($d
\leftrightarrow s$). This applies not only to the particles in the
process (e.g.\ $\pi^+ \leftrightarrow K^+$, $\bd \leftrightarrow \bs$,
etc.), but also to the individual diagrams involved. For any pair, one
can measure the two branching ratios and direct CP asymmetries in
order to obtain $X$ [Eq.~(\ref{Xdef})], and measure U-spin breaking.

\subsection{SU(3) pairs}
\label{2bodynonUspin}

U-spin pairs have been discussed at some length in
Refs.~\cite{GroUspin,NSL}. However, one can go further.  First, one
pair which is not included in the list in Sec.~\ref{2bodyUspin}, but
appears in Refs.~\cite{GroUspin,NSL}, is $\bs \to \pi^0 \kbar$ and
$\bd \to \pi^0K^0$.  The reason it is not included is that the two
decays are not related by U spin. There are a number of ways to see
this. First, $\pi^0 = (d{\bar d} - u{\bar u})/\sqrt{2}$, so that it
does not transform into itself under U spin. Second, one diagram that
contributes to $\bs \to \pi^0 \kbar$ is the penguin $P$, involving the
quark-level transition ${\bar b} \to {\bar d} d {\bar d}$. Under
U-spin reflection, this becomes ${\bar b} \to {\bar s} s {\bar s}$,
which does not contribute to $\bd \to \pi^0K^0$. What is going on is
the following: it is true that the amplitudes for $\bs \to \pi^0
\kbar$ and $\bd \to \pi^0K^0$ have the same diagrammatic decomposition
\cite{GHLR}, and so they satisfy Eq.~(\ref{Uspinrels}). However, the
diagrams assume isospin invariance in addition to U spin, so that the
symmetry is really flavor SU(3). Thus, $\bs \to \pi^0 \kbar$ and $\bd
\to \pi^0K^0$ is not a U-spin pair, but is in fact an SU(3) pair.

Second, it is standard to express the amplitudes for $B\to PP$ decays
in terms of diagrams \cite{GHLR}. Certain of these diagrams -- those
of annihilation- and exchange-type -- are expected to be smaller than
the others. If these diagrams are neglected, then there are additional
pairs of decays which satisfy Eq.~(\ref{Uspinrels}). These are not
related by U spin, but are instead related by SU(3). The complete list
of SU(3) pairs (which includes some U-spin pairs) is
\begin{itemize}

\item ($\bd \to \pi^+\pi^-$, $\bs \to \pi^+K^-$) and ($\bd \to
  \pi^-K^+$, $\bs \to K^+K^-$),

\item 
($\bd \to \pi^0\pi^0$, $\bs \to \pi^0 \kbar$, $\bs \to \eta_8 \kbar$)
  and ($\bd \to \pi^0K^0$, $\bd \to \eta_8 K^0$),

\item ($\bd \to K^0\kbar$, $B^+ \to K^+\kbar$, $\bd \to \pi^0 \eta_8$)
  and ($B^+ \to \pi^+K^0$, $\bs \to K^0\kbar$).

\end{itemize}
(Here, $\eta_8$ is a member of the octet of SU(3). The physical $\eta$
and $\eta'$ are linear combinations of $\eta_8$ and the SU(3) singlet,
$\eta_0$.) The decays in the first (second) parentheses are $\btod$
($\btos$) transitions. (Note that, depending on the pair, there may be
an additional factor (e.g.\ $\sqrt{2}$) in relating the $\btod$ and
$\btos$ decays.) From this list, we see that there are, in fact, 16
possible pairs of decays rather than the 5 of Sec.~\ref{2bodyUspin}.

If $(X-1)$ is obtained using a pair from Sec.~\ref{2bodyUspin}, then
U-spin breaking is measured. However, if an SU(3) pair is used, then
what is probed is not U-spin breaking, but rather SU(3) breaking.
Interestingly, we have data for a number of the decays in the above
list, so that it is possible to get $X$, and obtain an experimental
measurement of U-spin/SU(3) breaking in these decays. This is done in
Sec.~\ref{numerical}.

\subsection{\boldmath Estimates of $A_{s,d}$}
\label{estimates}

As described above, one can measure U-spin/SU(3) breaking through $X$.
This quantity involves the direct CP asymmetries $A_{d,s}$, which arise due to
the interference of two amplitudes with different weak and strong
phases. The maximal size of $A_{d,s}$ is roughly equal to the ratio of
the magnitudes of the two interfering amplitudes.

In two-body decays, the $\btos$ diagrams\footnote{The diagrams include
  the magnitudes of the associated CKM matrix elements.}  are expected
to obey the approximate hierarchy \cite{GHLR}
\bea
1 &:& |P'_{tc}| ~, \nn\\
{\bar\lambda} &:& |T'|~, |P'_{EW}| ~, \nn\\
{\bar\lambda}^2 &:& |C'|~, |P'_{uc}|~, |P^{\prime C}_{EW}| ~,
\label{btoshierarchy}
\eea
where ${\bar\lambda} \simeq 0.2$. Since all $\btos$ decays in the list
in Sec.~\ref{2bodynonUspin} receive contributions from $P'_{tc}$,
$A_s$ is sizeable ($\lsim O(\bar\lambda) \sim 20\%$) only if the decay
amplitude also includes $T'$. If there is no $T'$, but only $C'$ or
$P'_{uc}$, then $A_s$ is small ($\lsim O(\bar\lambda^2) \sim 5\%$). In
this case, the relative experimental error will necessarily be large,
which will then translate into a large error on $(X-1)$.

The expected approximate hierarchy\footnote{$C'$ and $C$ in
  Eqs.~(\ref{btoshierarchy}) and (\ref{btodhierarchy}) represent
  color-suppressed tree diagrams, and are not the parameters in
  Eq.~(\ref{fullamps}).} of the $\btod$ diagrams is \cite{GHLR}
\bea
1 &:& |T| ~, \nn\\
{\bar\lambda} &:& |C|~, |P_{tc}|~, |P_{uc}| ~, \nn\\
{\bar\lambda}^2 &:& |P_{EW}| ~, \nn\\
{\bar\lambda}^3 &:& |P^C_{EW}| ~.
\label{btodhierarchy}
\eea
Since all $\btod$ decays in the list in Sec.~\ref{2bodynonUspin}
receive penguin contributions, $A_d$ is always sizeable (at least
$\lsim O(\bar\lambda) \sim 20\%$).

Thus, the most promising pairs for measuring U-spin/SU(3) breaking are
those whose $\btos$ decay amplitude includes a $T'$. These are given
in the first entry in the list in Sec.~\ref{2bodynonUspin}.

There are two types of contributions to U-spin/SU(3) breaking --
factorizable and nonfactorizable. The factorizable effects depend
essentially on form factors and decay constants, and can be reliably
calculated. It has been shown that factorization holds well for
$T$/$T'$ diagrams \cite{BBNS}. Thus, for those decay pairs which
include these diagrams -- i.e.\ the most promising for measuring $X$
-- the ratio $|C'/C|$ is dominated by factorizable U-spin/SU(3)
breaking.

The U-spin relations $r' = r$ and $\delta'_r = \delta_r$ are not
affected by factorizable breaking effects\footnote{Decays such as
  $\bd\to\pi^0 K^0$ constitute an exception to this rule, as they can
  be factorized in two different ways. However, there are very few
  such decays.}, as the various form factors and decay constants
cancel \cite{Fleischer99,Fleischer}. On the other hand, they could be
altered by nonfactorizable effects, and these cannot be calculated
theoretically. Still, it is thought that nonfactorizable U-spin/SU(3)
breaking is not large, being higher-order in $1/m_b$. As we show
below, this can be checked experimentally through the measurement of
$(X-1)$.

\subsection{Numerical analysis}
\label{numerical}

The four quantities required for the measurement of $X$ are $B_{d,s}$
and $A_{d,s}$ [Eq.~(\ref{ABdefs})].  The $B_{d,s}$'s are related to
the branching ratios by
\beq
\tau_{(q)} p_{c(q)} B_{q} = 8 \pi \, m_{B(q)}^2 \mathcal{B}_{(q)} ~,
\eeq
where, for a $\bar b \to \bar q$ process ($q=d,s$), $\tau_{(q)}$ is
the $B$-meson lifetime, $p_{c(q)}$ is the momentum of the final-state
mesons in the $B$ rest frame, $m_{B(q)}$ is the rest mass of the $B$
meson, and $\mathcal{B}_{(q)}$ is the CP-averaged branching ratio. The
$A_{d,s}$'s are equal to $-C_{CP}$, where $C_{CP}$ is the direct CP
asymmetry in a given decay.

At present, there are five different pairs of two-body decays for
which we have the data required by the method of Sec.~\ref{Uspinbreak}
for measuring U-spin/SU(3) breaking:
\begin{enumerate}
	
\item[(1)] $\bd \to \pi^+ \pi^-$ and $\bd \to \pi^- K^+$,
	
\item[(2)] $\bs \to \pi^+ K^-$ and $\bd \to \pi^- K^+$,
	
\item[(3)] $B^+ \to K^+ \bar K^0$ and $B^+ \to \pi^+ K^0$,
	
\item[(4)] $\bd \to K^0 \bar K^0$ and $B^+ \to \pi^+ K^0$,

\item[(5)] $\bd \to \pi^0 \pi^0$ and $\bd \to \pi^0 K^0$.

\end{enumerate}
The current experimental values are given in Table \ref{tab:inputs1}.
The values of the $B$ masses and lifetimes can be found in
Ref.~\cite{pdg}.

\begin{table}[tbh]
\center
\begin{tabular}{|c|c|c|c|}
\hline
 Decay & $\mathcal{B}$ [$\times 10^6$] & $-C_{CP}$ & $p_c$ [MeV/$c$] \\
\hline
$\bd \to \pi^+ \pi^-$ & $5.16 \pm 0.22$ & $0.38 \pm 0.06$ & $2636$  \\
$\bd \to \pi^- K^+$ & $19.4 \pm 0.6$ & $-0.098^{+0.012}_{-0.011}$ & $2615$  \\
$\bs \to \pi^+ K^-$ & $5.0 \pm 1.1$ & $0.39 \pm 0.17$ & $2659$  \\
$B^+ \to K^+ \bar K^0$ & $1.36^{+0.29}_{-0.27}$ & $0.12^{+0.17}_{-0.18}$ & $2593$  \\
$B^+ \to \pi^+ K^0$ & $23.1 \pm 1.0$ & $0.009 \pm 0.025$ & $2614$  \\
$\bd \to K^0 \bar K^0$ & $0.96^{+0.21}_{-0.19}$ & $0.06 \pm 0.26$ & $2592$  \\
$\bd \to \pi^0 \pi^0$ & $1.55 \pm 0.19$ & $0.43^{+0.25}_{-0.24}$ & $2636$  \\
$\bd \to \pi^0 K^0$ (BaBar) & $10.1 \pm 0.6 \pm 0.4$ & $-0.13 \pm 0.13 \pm 0.03$ & $2615$  \\
$\bd \to \pi^0 K^0$ (Belle) & $8.7 \pm 0.5 \pm 0.6$ & $0.14 \pm 0.13 \pm 0.06$ &   \\
\hline
\end{tabular}
\caption{Input values for the experimental quantities \cite{pdg,hfag}.
  For asymmetric error bars, we take the average of both errors and
  assume a gaussian distribution.}
\label{tab:inputs1}
\end{table}

With these inputs, one can compute the value of $(X-1)$ obtained for
each of the five decay pairs using Eq.~(\ref{Xdef}). The results are
shown in Table \ref{tab:outputs}. Note that, as described in
Sec.~\ref{estimates}, the direct CP asymmetries in $B^+ \to \pi^+ K^0$
and $\bd \to \pi^0 K^0$ are expected to be quite small, leading to a
very large error on $(X-1)$. This is indeed what is found [pairs (3),
  (4) and (5)].

\begin{table}[tbh]
\center
\begin{tabular}{|c|c|}
\hline
Decay pair & $(X-1)$  \\
\hline
(1) & $-0.02 \pm 0.18$ \\
(2) & $-0.08 \pm 0.42$ \\
(3) & $-2.3 \pm 3.6$ \\
(4) & $-4 \pm 16$  \\
(5) & $1.0 \pm 2.1$ (BaBar) \\
& $-2.8 \pm 2.0$ (Belle) \\
\hline
\end{tabular}
\caption{Output values for the quantity $(X-1)$ for the five pairs of decays.}
\label{tab:outputs}
\end{table}

Finally, the decays $\bd \to \pi^+ \pi^-$ and $\bs \to K^+K^-$ form a
U-spin pair. From the updated QCD light-cone sum-rule calculation of
Ref.~\cite{LCSR2}, we have
\beq
\left\vert \frac{C'}{C} \right\vert_{fact} = \frac{f_K}{f_\pi}
\frac{f^+_{B_sK}(M_K^2)}{f^+_{B_d\pi}(M_\pi^2)} \left( \frac{M_{B_s}^2
  - M_K^2}{M_{B_d}^2 - M_\pi^2} \right) = 1.41^{+0.20}_{-0.11} ~.
\eeq
Here and below, we take $f^+(M_K^2) \simeq f^+(M_\pi^2) \simeq f^+(0)$
since the variation of the form factors over this range of $q^2$ falls
well within the errors of their calculation \cite{LCSR1}.  Thus, using
the data from Table \ref{tab:inputs1} and Eq.~(\ref{BdpiKBsKKdata})
below, we expect 
\bea
A_{CP}(\bs \to K^+K^-) & = & - \left\vert \frac{C'}{C}
\right\vert^2_{fact} A_{CP}(\bd \to \pi^+ \pi^-) \frac{\mathcal{B}(\bd
  \to \pi^+ \pi^-)}{\mathcal{B}(\bs \to K^+K^-)} \nn\\
& = & -0.16 \pm 0.05 ~.
\eea
Similarly, the decays $\bs \to \pi^+ K^-$ and $\bs \to K^+K^-$ form an
SU(3) pair, so that 
\bea
A_{CP}(\bs \to K^+K^-) & = & - \left\vert \frac{C'}{C}
\right\vert^2_{fact} A_{CP}(\bs \to \pi^+ K^-) \frac{\mathcal{B}(\bs \to \pi^+ K^-)}{\mathcal{B}(\bs \to K^+K^-)} \nn\\
& = & -0.12 \pm 0.06 ~,
\eea
where $|C'/C|_{fact} = f_K/f_\pi$. These predictions are in agreement
with one another and will be tested when $A_{CP}(\bs \to K^+K^-)$ is
measured.

\subsection{Measurement of nonfactorizable SU(3) breaking}
\label{nonfactmeasure}

The theoretical expression for $X$ is given in Eq.~(\ref{Xdef2}). As
discussed above, within factorization, only the ratio $|C'/C|$
contributes to $X$. Therefore, given an experimental measurement of
$X$ and a theoretical calculation of $|C'/C|_{fact}$, one can obtain
\beq
\frac{r' \sin{\delta'_r}}{r \sin{\delta_r}} = \left\vert \frac{C}{C'}\right\vert_{fact}^2 X
\eeq
and see whether it is consistent with 1 (small nonfactorizable
U-spin/SU(3) breaking).

For the first two pairs of the previous subsection, which yield
reasonably precise measurements of $X$, we have
\bea
{\hbox{pair (1)}} & : & \left\vert \frac{C'}{C} \right\vert_{fact} =
\frac{f_K}{f_\pi}
\frac{f^+_{B_d\pi}(M_K^2)}{f^+_{B_d\pi}(M_\pi^2)}
\left( \frac{M_{B_d}^2 - M_\pi^2}{M_{B_d}^2 - M_\pi^2} \right) \approx \frac{f_K}{f_\pi} = 1.20 ~, \nn\\
{\hbox{pair (2)}} & : & \left\vert \frac{C'}{C} \right\vert_{fact} =
\frac{f_K}{f_\pi}
\frac{f^+_{B_d\pi}(M_K^2)}{f^+_{B_sK}(M_\pi^2)}
\left( \frac{M_{B_d}^2 - M_\pi^2}{M_{B_s}^2 - M_K^2} \right) = 1.01^{+0.07}_{-0.15} ~.
\label{C'/C}
\eea
The ratio $f_K/f_\pi$ and the value in the second line are taken from
Ref.~\cite{LCSR2}. (We have neglected small errors in $f_K/f_\pi$.)
These give
\bea
{\hbox{pair (1)}} & : & \frac{r' \sin{\delta'_r}}{r \sin{\delta_r}} = 0.68 \pm 0.13 ~, \nn\\
{\hbox{pair (2)}} & : & \frac{r' \sin{\delta'_r}}{r \sin{\delta_r}} = 0.90 \pm 0.43 ~.
\eea

For pair (2), the theoretical prediction for factorizable U-spin
breaking is consistent with the experimental measurement of Table
\ref{tab:outputs}. However, for pair (1), there is a $2.5 \sigma$
deviation of the value of $|C/C'|_{fact}^2 X$ from 1. Now, as it is
just one data point, one cannot draw any firm conclusions -- it could
simply be a statistical fluctuation. However, it does hint at a large
nonfactorizable SU(3)-breaking correction. (Or, if one is certain that
such nonfactorizable effects are small, it could be suggestive of new
physics.) All of this illustrates the importance of measuring $X$
experimentally, and this in as many different decay pairs as possible.

This result does call into question any analysis which does not
include nonfactorizable corrections. However, it is straightforward to
take this into account. Within U-spin/SU(3) symmetry, the four
observables $B_{d,s}$ and $A_{d,s}$ are not independent. However, if
one allows U-spin/SU(3) breaking, this no longer holds. If one assumes
that $\delta'_r = \delta_r$, i.e.\ the phase is unaffected by the
breaking, and takes $|C'/C|$ from factorization, then nonfactorizable
U-spin/SU(3) breaking contributes only to $r'/r$. That is, there is
one additional theoretical parameter ($r'/r$), but there is one
additional measurement, so that the nonfactorizable breaking can be
obtained. This is essentially just the measurement of $X$.

Now, pair (1) is useful for another reason. As detailed previously, it
is not possible to obtain the theoretical parameters in the amplitudes
solely from the measurements of $B_{d,s}$ and $A_{d,s}$ -- additional
input is needed. This has been discussed for two of the U-spin
pairs. For $\bd \to \pi^+ \pi^-$ and $\bs \to K^+ K^-$, it has been
noted that $\gamma$ can be extracted through the additional
measurement of the indirect CP asymmetry in $\bd \to \pi^+ \pi^-$
\cite{Fleischer99,Fleischer}. Similarly, $\gamma$ can be obtained from $\bs \to
\pi^+ K^-$ and $\bd \to \pi^- K^+$ with the added information coming
from the measurement of the branching ratio of $B^+ \to \pi^+ K^0$
\cite{GRBsKpi}.

Both of these pairs appear in the list in Sec.~\ref{2bodyUspin}.
However, if one expands the symmetry from U spin to SU(3), they can be
combined, producing the pair $\bd \to \pi^+\pi^-$ and $\bd
\to\pi^-K^+$ (pair (1), in the list in Sec.~\ref{2bodynonUspin}).
$\gamma$ can then be extracted using the method of
Ref.~\cite{Fleischer99}, taking $B_d$, $A_d$ and $A^{CP}_{ind}$ from
$\bd \to \pi^+\pi^-$, and $B_s$ from $\bd \to \pi^-K^+$ instead of
$\bs \to K^+ K^-$. Since \cite{hfag, Tonelli} 
\bea
\mathcal{B}(\bd \to \pi^-K^+) &=& (19.4 \pm 0.6) \times 10^{-6} ~, \nn\\
\mathcal{B}(\bs \to K^+ K^-) &=& (23.9 \pm 3.9) \times 10^{-6} ~,
\label{BdpiKBsKKdata}
\eea
one sees that the first (experimental) error is smaller than the
second one. Thus, the error on $\gamma$ is also smaller.
Alternatively, suppose that the technique of Ref.~\cite{GRBsKpi} is
used, taking $B_s$ and $A_s$ from $\bd \to \pi^- K^+$, and $B_d$ from
$\bd \to \pi^+\pi^-$ instead of $\bs \to \pi^+ K^-$ (information from
$\mathcal{B}(B^+ \to \pi^+ K^0)$ is added).  The error on $\gamma$
will still be smaller since \cite{hfag}
\bea
\mathcal{B}(\bd \to \pi^+\pi^-) &=& (5.16 \pm 0.22) \times 10^{-6} ~, \nn\\
\mathcal{B}(\bs \to \pi^+ K^-) &=& (5.0 \pm 1.1) \times 10^{-6} ~.
\eea
The point is that the branching ratios of $\bd$ decays are measured
much more accurately than those of $\bs$ decays, so that the extracted
value of $\gamma$ is more precise if pair (1) is used, rather than
either U-spin pair.

In fact, this method was proposed many years ago, in 1995
\cite{GR95}. In this paper, information from both $A^{CP}_{ind}(\bd
\to \pi^+\pi^-)$ and $\mathcal{B}(B^+ \to \pi^+ K^0)$ is added
simultaneously. In addition, perfect SU(3) symmetry is not imposed, so
there are a total of 6 independent measurements. It is assumed that
$|C'/C| = f_K/f_\pi$ [Eq.~(\ref{C'/C})] and that $\delta'_r =
\delta_r$, but $r'$ and $r$ are left as independent. This means that
the amplitudes are written in terms of 4 hadronic theoretical
parameters and two weak phases. In Ref.~\cite{GR95}, it is argued that
both weak phases can be extracted. However, this method can be
modified: if we assume that $\beta$ is known from $A^{CP}_{ind}(\bd\to
J/\psi \ks)$, then we have the freedom to take $\delta'_r$ and
$\delta_r$ as independent. Now there are 6 equations in 6 unknowns
($C$, $r'$, $r$, $\delta'_r$, $\delta_r$, $\gamma$), so that one can
solve for the theoretical parameters (numerically, if necessary). This
analysis was partially performed in Ref.~\cite{GR2007}.  We must
stress here that no assumption about the size of nonfactorizable
effects in $r'/r$ and $\sin\delta'_r/\sin\delta_r$ is made here --
this information is taken from the experimental data.

\subsection{Other signals of SU(3) breaking}

There are pairs of decays whose amplitudes are equal at the quark
level, including CKM factors, under SU(3). At the meson level, the
processes are those within parentheses in the list in
Sec.~\ref{2bodynonUspin}. The amplitudes for the two decays can be
written
\beq
A_i = C_i^{(\prime)} \left[ \lambda_u^{(q)} + \lambda_t^{(q)} r_i^{(\prime)} e^{i \delta_{r,i}^{(\prime)}} \right] ~,
\eeq
where $i=1,2$ and $q=d,s$ (the hadronic parameters have primes for
$q=s$).  Assuming only factorizable SU(3) breaking, $r^{(\prime)}_1 =
r^{(\prime)}_2$ and $\delta_{r,1}^{(\prime)} =
\delta_{r,2}^{(\prime)}$. We therefore expect the branching ratios
and direct CP asymmetries for the two decays to satisfy
\bea
\mathcal{B}_2 & = & \left\vert \frac{C^{(\prime)}_2}{C^{(\prime)}_1} \right\vert_{fact}^2
\mathcal{B}_1 ~, \nn\\
A_{CP,2} & = & A_{CP,1} ~.
\eea
(We neglect any mass and lifetime differences between the two decaying
$B$ mesons.)  Any deviation from these relations is a sign of
nonfactorizable SU(3) breaking.

The pairs or amplitude relations are (all experimental data is taken
from Ref.~\cite{hfag}):
\begin{enumerate}

\item $\bd \to \pi^-K^+$ and $\bs \to K^+K^-$:
\beq
\left\vert \frac{C'_1}{C'_2} \right\vert_{fact} =
\frac{f^+_{B_d\pi}(M_K^2)}{f^+_{B_sK}(M_K^2)}
\left( \frac{M_{B_d}^2 - M_\pi^2}{M_{B_s}^2 - M_K^2} \right) = 0.85^{+0.07}_{-0.12} ~.
\label{C1/C2}
\eeq
(This is based on the results of Ref.~\cite{LCSR2}.) The data for the
branching ratios for these decays are given in
Eq.~(\ref{BdpiKBsKKdata}). We expect
\beq
\left\vert \frac{C'_2}{C'_1} \right\vert_{fact}^2 \frac{\mathcal{B}(\bd \to \pi^-K^+)}{\mathcal{B}(\bs \to K^+ K^-)}
\eeq
to be consistent with 1. It equals $1.12 \pm 0.26$, so there is no
evidence of nonfactorizable SU(3) breaking.

We also expect that
\beq
A_{CP}(\bs \to K^+K^-) = A_{CP}(\bd \to \pi^-K^+) = -0.098^{+0.012}_{-0.011} ~.
\eeq

\item $\bd \to \pi^+\pi^-$ and $\bs \to \pi^+K^-$: here, $\left\vert
  C_1/C_2 \right\vert_{fact} = 0.85^{+0.07}_{-0.12}$, as in
  Eq.~(\ref{C1/C2}). The experimental data is: $\mathcal{B}(\bd \to
  \pi^+\pi^-) = (5.16 \pm 0.22) \times 10^{-6}$, $\mathcal{B}(\bs \to
  \pi^+K^-) = (5.0 \pm 1.1) \times 10^{-6}$. We expect
\beq
\left\vert \frac{C_2}{C_1} \right\vert_{fact}^2 \frac{\mathcal{B}(\bd \to \pi^+\pi^-)}{\mathcal{B}(\bs \to \pi^+K^-)}
\eeq
to be consistent with 1. It equals $1.43 \pm 0.40$. We also expect the
direct CP asymmetries to be equal. It is found that $A_{CP}(\bd \to
\pi^+\pi^-) = 0.38 \pm 0.06$, $A_{CP}(\bs \to \pi^+K^-) = 0.39 \pm
0.17$, which are in good agreement with one another.  We therefore
conclude that there is no evidence for nonfactorizable SU(3) breaking
in this decay pair.

\item $A(\bd \to \pi^0 K^0) = \sqrt{3} A(\bd \to \eta_8 K^0)$: we
  expect $\mathcal{B}(\bd \to \pi^0 K^0) = \left\vert C'_1/C'_2
  \right\vert_{fact}^2 \\ 3 \mathcal{B}(\bd \to \eta_8 K^0)$ and
  $A_{CP}(\bd \to \pi^0 K^0) = A_{CP}(\bd \to \eta_8 K^0)$.

\item $A(\bd \to \pi^0 \pi^0) = A(\bs \to \pi^0 \kbar) = \sqrt{3}
  A(\bs \to \eta_8 \kbar)$: this leads to the prediction that
  $A_{CP}(\bs \to \pi^0 \kbar) = A_{CP}(\bs \to \eta_8 \kbar) =
  0.43^{+0.25}_{-0.24}$. Also, we expect that $\mathcal{B}(\bs \to
  \pi^0 \kbar) = (1.55 \pm 0.16) \times 10^{-6}$, $\mathcal{B}(\bs \to
  \eta_8 \kbar) = (0.52 \pm 0.05) \times 10^{-6}$, modulo factorizable
  SU(3) corrections.

\item $A(B^+ \to \pi^+ K^0) = A(\bs \to K^0 \kbar)$, so that the
  direct CP asymmetries are expected to be equal for these decays, and
  we expect $\mathcal{B}(B^+ \to \pi^+ K^0) = \left\vert C'_1/C'_2
  \right\vert_{fact}^2 \mathcal{B}(\bs \to K^0 \kbar)$.

\item $A(\bd \to K^0 \kbar) = A(B^+ \to K^+ \kbar) = \sqrt{3} A(\bd
  \to \pi^0 \eta_8)$: we expect the direct CP asymmetries for these
  three decays to be equal.  Also, we expect that $\mathcal{B}(\bd \to
  K^0 \kbar) = \mathcal{B}(B^+ \to K^+ \kbar) = 3 \mathcal{B}(\bd \to
  \pi^0 \eta_8)$, modulo factorizable SU(3) corrections.

\end{enumerate}
Note: it would not be a surprise to see evidence of significant
nonfactorizable effects in the decays in items 4-6, as these are
dominated by diagrams for which factorization is not expected to hold.

\section{Three-Body Decays}

We now turn to $B\to PPP$ decays. In the past, such decays were little
studied -- most of the theoretical work looking at clean methods for
obtaining the weak phases focused on two-body $B$ decays.  This is
essentially for two reasons: (i) final states such as $\psi \ks$,
$\pi^+\pi^-$, etc.\ are CP eigenstates, and (ii) when there is a
second decay amplitude, with a different weak phase, it has been
possible to find methods to remove this ``pollution,'' and cleanly
extract weak-phase information.

Things are not the same in the case of three-body $B$ decays. First,
because there are three particles, final states such as $\ks
\pi^+\pi^-$ are not CP eigenstates -- the value of its CP depends on
whether the relative $\pi^+\pi^-$ angular momentum is even or odd. And
second, even if it were possible to distinguish the states of CP $+$
and $-$, one still has the problem of removing the pollution due to
additional decay amplitudes. For these reasons, the conventional
wisdom has been that it is not possible to obtain clean weak-phase
information from three-body decays.

Recently, it was shown that, by doing a diagrammatic analysis, one can
address these two problems \cite{diagramspaper}.  First, a Dalitz-plot
analysis can be used to experimentally separate the CP $+$ and $-$
final states. Second, one can often remove the pollution of additional
diagrams and cleanly measure the CP phases. In Ref.~\cite{Kpipipaper},
the procedure for extracting $\gamma$ from $B\to K\pi\pi$ decays was
described in detail. Thus, in fact, it {\it is} possible to use
three-body decays to obtain weak-phase information and search for new
physics.

In this paper, the goal is to find pairs of $\btod$ and $\btos$ decays
which satisfy Eq.~(\ref{Uspinrels}) and permit the measurement of $X$.
As we will see, in order to do this with three-body decays, the
diagrammatic decomposition of Ref.~\cite{diagramspaper} is necessary.

\subsection{U-spin pairs}

As with $B\to PP$ decays (Sec.~\ref{2bodyUspin}), we look for pairs of
$\btos$ and $\btod$ decays that are related by U-spin reflection. We
find that there are seven such pairs of three-body decays:
\begin{enumerate}
	
\item $\bs \to K^+ K^- \kbar$ and $\bd \to K^0 \pi^+ \pi^-$, 

\item $\bs \to \kbar \pi^+\pi^-$ and $\bd \to K^+ K^0 K^-$,
	
\item $\bd \to K^0 K^- \pi^+$ and $\bs \to K^+\kbar \pi^-$,
	
\item $\bd \to K^+ \kbar \pi^-$ and $\bs\to K^0K^-\pi^+$,
	
\item $B^+ \to \pi^+ \pi^+ \pi^-$ and $B^+ \to K^+ K^+ K^-$,
	
\item $B^+ \to K^+ K^- \pi^+$ and $B^+ \to K^+ \pi^+ \pi^-$,
	
\item $\bs \to \kbar \kbar K^0$ and $\bd \to K^0 K^0 \kbar$.

\end{enumerate}
In order to show that these pairs do indeed satisfy
Eq.~(\ref{Uspinrels}), one has to compare the amplitudes of the decays
within a pair.  

Under U spin, the $d$ and $s$ quarks are in a doublet, as are ${\bar
  s}$ and $-{\bar d}$. Thus, $K^+$ and $\pi^+$, and $K^-$ and $\pi^-$,
are considered to be identical particles. We therefore see that the
final states of pairs 1-4 contain no identical particles. One can
straightforwardly compare the amplitudes of the decays within these
pairs. We refer to Ref.~\cite{diagramspaper} for a description of the
diagrams; here we label each diagram $D$ by an index $q$ ($q=u,d,s$)
denoting the flavor of the quark pair ``popped'' from the
vacuum. Under isospin symmetry, $D_u = D_d$, under U spin, $D_d =
D_s$, and under full SU(3), $D_u = D_d = D_s$. We have:

\noindent
pair 1:
\bea
A(\bs \to K^+ K^- \kbar) &=& - T_{1,s} e^{i\gamma} - C_{1,s}
e^{i\gamma} - {\hat P}_{a;uc} e^{i\gamma} \nn\\
&& \hskip-1.5truein -~{\hat P}_{a;tc} e^{-i\beta} - \frac23 P_{EW1,s}
e^{-i\beta} + \frac13 P_{EW1,u} e^{-i\beta} - \frac23 P_{EW1,s}^C
e^{-i\beta} + \frac13 P_{EW2,u}^C e^{-i\beta} ~,\nn\\
A(\bd \to K^0 \pi^+ \pi^-) &=& -T'_{1,d} e^{i\gamma}-C'_{1,d}
e^{i\gamma}-{\tilde P}'_{a;uc} e^{i\gamma}  \nn\\
&& \hskip-0.5truecm +~{\tilde P}'_{a;tc} + \frac23 P'_{EW1,d} - \frac13 P'_{EW1,u}
+ \frac23 P^{\prime C}_{EW1,d} - \frac13 P^{\prime C}_{EW2,u} ~,
\eea
pair 2:
\bea
A(\bs \to \bar K^0 \pi^+ \pi^-) &=& -T_{2,d} e^{i\gamma} - C_{1,d}
e^{i\gamma} - {\tilde P}_{b;uc} e^{i\gamma} \nn\\
&& \hskip-1.5truein -~{\tilde P}_{b;tc} e^{-i\beta} - \frac23 P_{EW1,d} e^{-i\beta} +~\frac13 P_{EW1,u} e^{-i\beta} + \frac13
P_{EW1,u}^C e^{-i\beta} - \frac23 P_{EW2,d}^C e^{-i\beta} ~,\nn\\
A(\bd \to K^+ K^0 K^-) &=& -T'_{2,s} e^{i\gamma}-C'_{1,s} e^{i\gamma}
-{\hat P}'_{b;uc} e^{i\gamma}+ {\hat P}'_{b;tc} \nn\\
&& \hskip0.5truecm +~\frac23 P'_{EW1,s} - \frac13 P'_{EW1,u} - \frac13
P^{\prime C}_{EW1,u} + \frac23 P^{\prime C}_{EW2,s} ~,
\eea
pair 3:
\bea
A(\bd \to K^0 K^- \pi^+) &=& - T_{2,s} e^{i \gamma} -{\hat P}_{b;uc} e^{i \gamma}
 \nn\\
&& -~{\hat P}_{b;tc} e^{-i\beta} + \frac13 P^C_{EW1,u}e^{-i\beta} - \frac23 P^C_{EW2,s} e^{-i\beta}~, \nn\\
A(\bs \to K^+\kbar \pi^-) &=& - T'_{2,d} e^{i \gamma} -{\tilde P}'_{b;uc}
e^{i \gamma} + {\tilde P}'_{b;tc} - \frac13 P_{EW1,u}^{'C} + \frac23
P_{EW2,d}^ {'C} ~, 
\eea
pair 4:
\bea
A(\bd \to K^+\kbar\pi^-) & = & -T_{1,s} e^{i\gamma} - {\hat P}_{a;uc} e^{i\gamma} \nn\\
&&  -~{\hat P}_{a;tc} e^{-i\beta} -~\frac23 P^C_{EW1,s} e^{-i\beta} + \frac13 P^C_{EW2,u} e^{-i\beta}~, \nn\\
A(\bs\to K^0K^-\pi^+) &=& - T'_{1,d} e^{i \gamma} -{\tilde P}'_{a;uc} e^{i
  \gamma} + {\tilde P}'_{a;tc} + \frac23 P_{EW1,d}^{'C} - \frac13
P_{EW2,u}^{'C} ~, 
\eea
where 
\bea
{\tilde P}_a \equiv P_{1,d} +P_{2,u} & ~~,~~~~ & {\tilde P}_b \equiv
P_{1,u} +P_{2,d} ~, \nn\\
{\hat P}_a \equiv P_{1,s} + P_{2,u} & ~~,~~~~ & {\hat P}_b \equiv
P_{1,u} + P_{2,s} ~.
\label{Pdefs1}
\eea
For $\btod$ transitions, the diagrams are written without primes; for
$\btos$ transitions, they are written with primes.  (The overall signs
of the amplitudes assume ${\bar u}$ is negative, as with isospin. If
one takes ${\bar d}$ to be negative, as with U spin, one may obtain a
different overall sign. But the physics does not change.)

There are two truly identical particles in the final states in pair 5
($\pi^+$ in $B^+ \to \pi^+ \pi^+ \pi^-$ and $K^+$ in $B^+ \to K^+ K^+
K^-$), so the overall wavefunction must be symmetric with respect to
the exchange of these two particles: 
\bea
A(B^+\to \pi^+\pi^+\pi^-)_{sym} &=& -T_{2,d}
e^{i\gamma} - C_{1,d} e^{i\gamma} - {\tilde P}_{b;uc} e^{i\gamma} \nn\\
&& \hskip-1.5truein -~{\tilde P}_{b;tc} e^{-i\beta} - \frac23 P_{EW1,d} e^{-i\beta} + \frac13 P_{EW1,u} e^{-i\beta} + \frac13 P^{C}_{EW1,u} e^{-i\beta} - \frac23
P^{C}_{EW2,d} e^{-i\beta} ~, \nn\\
A(B^+ \to K^+ K^+ K^-)_{sym} &=& -T'_{2,s} e^{i\gamma} -C'_{1,s}
e^{i\gamma} -{\hat P}'_{b;uc} e^{i\gamma} \nn\\
&& \hskip-0.8truein +~{\hat P}'_{b;tc} + \frac23 P'_{EW1,s} - \frac13 P'_{EW1,u} - \frac13 P^{\prime C}_{EW1,u} + \frac23
P^{\prime C}_{EW2,s} ~.
\eea
The penguin diagrams are defined in Eq.~(\ref{Pdefs1}).

The final states of pair 6 contain the identical particles (under
U spin) $K^+$ and $\pi^+$.  The overall wavefunction of the final
$K^+\pi^+$ pair must be symmetrized with respect to the exchange of
these two particles. If the relative angular momentum is even (odd),
the U-spin state must be symmetric (antisymmetric):
\bea
A(B^+ \to K^+K^-\pi^+)_{sym} &=& -T_{2,s} e^{i\gamma} - C_{1,s}
e^{i\gamma} - {\hat P}_{b;uc} e^{i\gamma} - {\hat P}_{b;tc} e^{-i\beta} \nn\\ 
&& \hskip-2.2truecm +~\frac13 P_{EW1,u} e^{-i\beta} -\frac23 P_{EW1,s} e^{-i\beta} + \frac13
P^C_{EW1,u} e^{-i\beta} - \frac23 P^C_{EW2,s} e^{-i\beta} ~, \nn\\
A(B^+ \to K^+K^-\pi^+)_{anti} &=& T_{2,s} e^{i\gamma} + C_{1,s}
e^{i\gamma} + {\hat P}_{b;uc} e^{i\gamma} + {\hat P}_{b;tc} e^{-i\beta} \nn\\ 
&& \hskip-2.2truecm +~\frac13 P_{EW1,u} e^{-i\beta} -\frac23 P_{EW1,s} e^{-i\beta} + \frac13
P^C_{EW1,u} e^{-i\beta} - \frac23 P^C_{EW2,s} e^{-i\beta} ~, \nn\\
A(B^+ \to K^+\pi^+\pi^-)_{sym} &=& -T'_{2,d} e^{i\gamma}-C'_{1,d} e^{i\gamma}-{\tilde P}'_{b;uc} e^{i\gamma}+ {\tilde P}'_{b;tc} \nn\\
&& \hskip0.2truecm -~\frac13 P'_{EW1,u} + \frac23 P'_{EW1,d} - \frac13 P^{\prime C}_{EW1,u} + \frac23 P^{\prime C}_{EW2,d} ~, \nn\\
A(B^+ \to K^+\pi^+\pi^-)_{anti} &=& -T'_{2,d} e^{i\gamma}-C'_{1,d} e^{i\gamma}-{\tilde P}'_{b;uc} e^{i\gamma}+ {\tilde P}'_{b;tc} \nn\\
&& \hskip0.2truecm +~\frac13 P'_{EW1,u} - \frac23 P'_{EW1,d} + \frac13 P^{\prime C}_{EW1,u} - \frac23 P^{\prime C}_{EW2,d} ~, 
\eea
where, for the antisymmetric amplitudes, diagrams with the $K^+$ above
(below) the $\pi^+$ are multiplied by $+1$ ($-1$). The penguin
diagrams are defined in Eq.~(\ref{Pdefs1}).

Both the $\kbar$ and $K^0$ are contained in a U-spin triplet, and so
these are considered as identical particles.  Thus, the final states
of the decays in pair 7 contain three identical particles and the
group $S_3$ must be used to describe their permutations. Fortunately,
for these decays, the situation is less complicated.  For $K^0 K^0
\kbar$, in any diagram, the position of the $\kbar$ cannot change, so
that only exchanges of the two $K^0$'s need be considered. Things are
similar for $\kbar \kbar K^0$.  Thus, in order to show that these
decays do indeed form a pair which respects Eq.~(\ref{amps}), it is
sufficient to examine the amplitudes which are symmetric in the
exchange of the two truly identical particles. We have
\bea
A(\bs \to \kbar \kbar K^0)_{sym} &=& {\mathcal P}_{a;uc} e^{i\gamma} +
{\mathcal P}_{a;tc} e^{-i\beta} - \frac13 P_{EW1,s} e^{-i\beta} - \frac13
P_{EW1,d} e^{-i\beta} \nn\\
&& \hskip0.8truecm   -~\frac13 P_{EW1,s}^C e^{-i\beta} - \frac13 P_{EW2,d}^C e^{-i\beta}
~, \nn\\
A(\bd \to K^0 K^0 \kbar)_{sym} &=& {\mathcal P}'_{b;uc} e^{i\gamma}- {\mathcal P}'_{b;tc} + \frac13 P'_{EW1,s} + \frac13 P'_{EW1,d} \nn\\
&& \hskip0.8truecm +~\frac13 P^{\prime C}_{EW1,d} + \frac13
P^{\prime C}_{EW2,s}  ~,
\eea
where 
\beq
{\mathcal P}_a \equiv P_{1,s} +P_{2,d} ~~,~~~~ {\mathcal P}_b \equiv P_{1,d} +P_{2,s} ~.
\label{Pdefs2}
\eeq

Now, under U spin, primed diagrams are equal to unprimed diagrams with
the exchange $d \leftrightarrow s$, i.e.\ they differ only by
$\lambda_p^{(d)} \leftrightarrow \lambda_p^{(s)}$.  Thus, $D'_s \sim
D_d$, $D'_d \sim D_s$, $D'_u \sim D_u$, ${\tilde P}'_a \sim {\hat
  P}_a$, ${\tilde P}_a \sim {\hat P}'_a$, ${\tilde P}'_b \sim {\hat
  P}_b$, ${\tilde P}_b \sim {\hat P}'_b$, and ${\mathcal P}_a \sim
{\mathcal P}_b$. We therefore see that (almost all) the amplitudes for
the $\btod$ and $\btos$ decays in pairs 1-7 have the same form, modulo
CKM factors (recall that the $\btos$ amplitudes include the minus sign
from $V_{tb}^* V_{ts}$ [$P'_{tc}$ and EWP's]).  The single exception
is $A(B^+ \to K^+K^-\pi^+)_{anti}$ and $A(B^+ \to
K^+\pi^+\pi^-)_{anti}$ in pair 6. Here, recall that the contribution
of a diagram is positive (negative) if the $K^+$ is above (below) the
$\pi^+$ in that diagram. However, since the U-spin transformation
switches $K^+ \leftrightarrow \pi^+$, we expect the antisymmetric
amplitudes to have a relative $-$ sign, and this is indeed what is
found.  We therefore see that, in all pairs, the amplitudes of the
$\btod$ and $\btos$ decays respect Eq.~(\ref{Uspinrels}), so that $(X-1)$
(U-spin breaking) can be measured using these processes.

Previously, in discussing two-body decays, we noted that the
U-spin/SU(3) corrections could be separated into two types --
factorizable and nonfactorizable -- and that the factorizable
corrections could be reliably calculated. As such, the measurement of
$X$ can be translated into a determination of the nonfactorizable
corrections. In principle, this can be applied to three-body decays.
In practice, however, things are more complicated. In particular,
while the $T^{(\prime)}$ diagram in two-body decays is, within
factorization, proportional to the product of a decay constant and a
form factor, in three-body decays, new structures appear.  The
$T_1^{(\prime)}$ diagram is proportional to the product of a $\langle
2 ~{\rm particles} | (V-A) | 0 \rangle$ matrix element and a form
factor, and the $T_2^{(\prime)}$ diagram is proportional to the
product of a decay constant and a $\langle 2 ~{\rm particles} | (V-A)
| B \rangle$ matrix element. To date, there have been no definitive
calculations of these matrix elements. They have been studied in
Ref.~\cite{3fact}, but more work is clearly needed. 

To this end, the measurement of $X$ can help. Given that
nonfactorizable U-spin/SU(3) breaking is expected to be subdominant
compared to factorizable breaking, $X$ as measured in the above decay
pairs can be considered to be a factorizable correction (especially
pairs 1-6, which have $T_1^{(\prime)}/T_2^{(\prime)}$
contributions). The knowledge of the precise values of such
factorizable effects will guide the calculation of the new matrix
elements.

Finally, a natural question is whether clean weak-phase
information can be extracted from these decays. For example, the pair
$\bs \to K^+ K^- \kbar$ and $\bd \to K^0 \pi^+ \pi^-$ is the
three-body equivalent of $\bd \to \pi^+ \pi^-$ and $\bs \to K^+
K^-$. Can one adapt the method of Ref.~\cite{Fleischer99} to obtain
$\gamma$?  Unfortunately, the answer is no. In two-body decays,
additional information is provided by the measurement of the indirect
CP asymmetry in $\bd \to \pi^+ \pi^-$. Here, however, because $\bd \to
K^0 \pi^+ \pi^-$ is a three-body decay, the relative $\pi^+\pi^-$
angular momentum is not fixed, and so the final state is not a CP
eigenstate. Thus, the measurement of the indirect CP asymmetry in this
decay does not give clean information. The situation is the same for
the second pair, $\bs \to \kbar \pi^+\pi^-$ and $\bd \to K^+ K^0 K^-$.

In a similar vein, $\bd \to K^0 K^- \pi^+$ and $\bs \to K^+\kbar
\pi^-$ is the three-body equivalent of $\bs \to \pi^+ K^-$ and $\bd
\to \pi^- K^+$. Can the method of Ref.~\cite{GRBsKpi}, in which
additional information comes from $\mathcal{B}(B^+ \to \pi^+ K^0)$, be
adapted to this situation? Unfortunately, here too the answer is
no. Unlike the two-body situation, here there is no other three-body
decay which provides the appropriate additional information. This
holds as well for pairs 4-6.

On the other hand, pair 7, $\bs \to \kbar \kbar K^0$ and $\bd \to K^0
K^0 \kbar$ is intriguing. The key point here is that, because there
are truly identical particles in the final state, their relative
angular momentum is even, and so the final state {\it is} a CP eigenstate.
Now, the diagram contributing to the $e^{i\gamma}$ piece of the
$\btos$ amplitude is ${\mathcal P}'_{b;uc}$. In Sec.~\ref{estimates},
we noted that $|P'_{uc}|$ is expected to be small in two-body decays,
and so a direct CP asymmetry which is proportional to this diagram
will also be small. If the same property holds in three-body decays,
the measurement of the indirect CP asymmetry in the pure-penguin decay
$\bs \to \kbar \kbar K^0$ cleanly probes the $\bs$-$\bsbar$ mixing
phase (experimentally, this might be easier than performing the
angular analysis in $\bs\to J/\psi \phi$, which is presently
done). However, if ${\mathcal P}'_{b;uc}$ is not small, as could
happen if there are significant rescattering effects, then $A_s$ is
not negligible, and the method of Ref.~\cite{Fleischer99} can be
applied to this pair to obtain $\gamma$. Here, U-spin symmetry is
assumed, but, as noted above, it is possible to measure $X$, which
gives the size of U-spin breaking.

\subsection{SU(3) pairs}
\label{3bodynonUspin}

Unlike two-body decays, with three-body decays one cannot obtain
additional pairs satisfying Eq.~(\ref{Uspinrels}) by simply neglecting
annihilation- and exchange-type diagrams. However, there is another
possibility. If, as in the two-body case, one takes isospin into
account in addition to U-spin symmetry, one effectively assumes full
flavor SU(3) symmetry.  Under this symmetry, $\pi$'s and $K$'s are
identical particles, so that the final state in all decays contains
three identical particles. In this case, the six permutations of these
particles (the group $S_3$) must be considered.  This was analyzed in
Ref.~\cite{diagramspaper}.  For a given decay, there are six
possibilities for the $S_3$ state of the three particles: a totally
symmetric state $\ket{S}$, a totally antisymmetric state $\ket{A}$, or
one of four mixed states $\ket{M_i}$ ($i=1$-4). The states are defined
as follows. The final-state particles are numbered 1, 2, 3, so that
the six possible orders are 123, 132, 312, 321, 231, 213. Under $S_3$,
\bea
\ket{S} &\equiv& \frac{1}{\sqrt{6}} \left( \ket{123} + \ket{132} + \ket{312} + \ket{321} + \ket{231} + \ket{213} \right)~,\nn\\
\ket{M_1} &\equiv& \frac{1}{\sqrt{12}} \left( 2\ket{123} + 2\ket{132} - \ket{312} - \ket{321} - \ket{231} - \ket{213} \right)~,\nn\\
\ket{M_2} &\equiv& \frac{1}{\sqrt{4}} \left( \ket{312} - \ket{321} - \ket{231} + \ket{213} \right)~,\nn\\
\ket{M_3} &\equiv& \frac{1}{\sqrt{4}} \left( -\ket{312} - \ket{321} + \ket{231} + \ket{213} \right)~,\nn\\
\ket{M_4} &\equiv& \frac{1}{\sqrt{12}} \left( 2\ket{123} - 2\ket{132} - \ket{312} + \ket{321} - \ket{231} + \ket{213} \right)~,\nn\\
\ket{A} &\equiv& \frac{1}{\sqrt{6}} \left( \ket{123} - \ket{132} + \ket{312} - \ket{321} + \ket{231} - \ket{213} \right)~.
\label{SU3states}
\eea

One can show that certain pairs of decays, related by SU(3) and not by
U spin, satisfy Eq.~(\ref{amps}), but only for the state $\ket{S}$ (in
most cases). This applies to the following SU(3) pairs\footnote{Note:
  this list includes some U-spin pairs. These pairs are related for
  all $S_3$ states.} (as is standard, we neglect annihilation- and
exchange-type diagrams): 
\begin{itemize}
	
\item ($B^+ \to \pi^+ K^- K^+$, $B^+ \to \pi^+ \pi^0 \pi^0$, $B^+ \to
  \pi^+ \pi^+ \pi^-$, $\bs \to \kbar \pi^+\pi^-$) and ($\bd \to
  K^+K^-K^0$, $B^+ \to K^+K^+K^-$, $B^+ \to K^+ \pi^+\pi^-$),
	
\item ($B^+ \to K^+ \kbar \pi^0$, $B^+ \to \kbar K^+ \eta_8$) and ($B^+ \to K^0 \pi^+ \pi^0$, $\bd
  \to K^+ \pi^- \pi^0$, $B^+ \to K^0 \pi^+ \eta_8$),
	
\item $\bs \to \kbar \kbar K^0$ and ($B^+ \to K^+ K^0 \kbar$, $\bd \to
  K^0 K^0 \kbar$),
	
\item $\bd \to \pi^0 \pi^0 \pi^0$ and $\bd \to K^0 \pi^0 \pi^0$,

\item ($\bd \to K^- K^+ \pi^0$, $\bd \to K^- K^+ \eta_8$) and ($\bs \to \pi^0 \pi^0 \eta_8$, $\bs \to \pi^+ \pi^- \eta_8$),

\item $\bs \to \kbar \pi^0 \pi^0$ and $\bd \to K^0 \pi^0 \pi^0$,

\item $\bs \to \kbar \pi^0 \eta_8$ and $\bd \to K^0 \pi^0 \eta_8$,

\item $\bs \to \kbar \eta_8 \eta_8$ and $\bd \to K^0 \eta_8 \eta_8$.

\end{itemize}
The decays in the first (second) parentheses are $\btod$ ($\btos$)
transitions. 

In order to establish which states are the same (modulo CKM factors)
for the decays within a pair, one writes the amplitudes for each decay
in terms of diagrams, noting the order of the final-state particles
for each diagram. It is this order which determines which $S_3$ states
are common to both decays. The state $\ket{S}$ is symmetric in all
possible orders. Thus, as long as the two amplitudes are comprised of
the same diagrams, the final-state order is unimportant, and the two
decays are related by SU(3) for $\ket{S}$. For $\ket{A}$, if the first
decay amplitude contains the diagram $D$ with the order $ijk$, the
second decay amplitude must contain $D$ with a cyclic permutation of
$ijk$, or $-D$ with a anticyclic permutation of $ijk$.

The mixed states are more complicated. The six elements of $S_3$ are:
$I$ (identity), $P_{12}$ (exchanges particles 1 and 2), $P_{13}$
(exchanges particles 1 and 3), $P_{23}$ (exchanges particles 2 and 3),
$P_{cyclic}$ (cyclic permutation of particle numbers, i.e.\ $1\to 2$,
$2\to 3$, $3\to 1$), $P_{anticyclic}$ (anticyclic permutation of
particle numbers, i.e.\ $1\to 3$, $2\to 1$, $3\to 2$). The point is
that, under the group transformations, $\ket{M_1}$ and $\ket{M_3}$
transform among themselves. Writing
\beq
\ket{M_1} \equiv \left( \matrix{1 \cr 0} \right) ~~,~~~~ 
\ket{M_3} \equiv \left( \matrix{0 \cr 1} \right) ~~,
\eeq
we can represent each group element by a $2\times 2$ matrix:
\bea
& I = \left( \matrix{1 & 0 \cr 0 & 1} \right) ~,~~
P_{12} = \left( \matrix{-\frac12 & \frac{\sqrt{3}}{2} \cr \frac{\sqrt{3}}{2} & \frac12} \right) ~,~~
P_{13} = \left( \matrix{-\frac12 & -\frac{\sqrt{3}}{2} \cr -\frac{\sqrt{3}}{2} & \frac12} \right) ~, & \nn\\
& P_{23} = \left( \matrix{1 & 0 \cr 0 & -1} \right) ~,~~
P_{cyclic} = \left( \matrix{-\frac12 & \frac{\sqrt{3}}{2} \cr -\frac{\sqrt{3}}{2} & -\frac12} \right) ~,~~
P_{anticyclic} = \left( \matrix{-\frac12 & -\frac{\sqrt{3}}{2} \cr \frac{\sqrt{3}}{2} & -\frac12} \right) ~. &
\label{matrices}
\eea
Similarly, if we write
\beq
\ket{M_2} \equiv \left( \matrix{1 \cr 0} \right) ~~,~~~~ 
\ket{M_4} \equiv \left( \matrix{0 \cr 1} \right) ~~,
\eeq
the $S_3$ matrices take the same form, showing that $\ket{M_2}$ and
$\ket{M_4}$ also transform among themselves.

{}From the above matrices, we see that the first rows of the matrices
are the same for ($I$, $P_{23}$), ($P_{12}$, $P_{cyclic}$) and
($P_{13}$, $P_{anticyclic}$). This indicates that the symmetric mixed
states ($\ket{M_1}$ and $\ket{M_2}$) are the same for the two decays
if the particle orders for a given diagram are [(123) or (132)],
[(213) or (231)], or [(321) or (312)]. For the antisymmetric mixed
states ($\ket{M_3}$ and $\ket{M_4}$), things are the same, except that
there is an additional minus sign if the particle order is anticyclic
(this can be seen from the second rows of the matrices).

To demonstrate how this works, we present several examples. First,
consider the decays $B^+ \to \pi^+ K^- K^+$ and $\bd \to K^+K^-K^0$.
For $B^+ \to \pi^+ K^- K^+$ we take particle 1 as $\pi^+$, particle 2
as $K^-$, and particle 3 as $K^+$. The amplitude is
\bea
A(B^+ \to \pi^+ K^- K^+) &=& -T_{2,s}e^{i\gamma}(123) -
C_{1,s}e^{i\gamma}(132) - {\hat P}_{b;uc} e^{i\gamma} (123) \nn\\
&& \hskip-0.8truecm -~{\hat P}_{b;tc} e^{-i\beta} (123) + \frac13 P_{EW1,u}e^{-i\beta}(231) -\frac23
P_{EW1,s}e^{-i\beta}(321) \nn\\
&& \hskip-0.8truecm +~\frac13 P^C_{EW1,u}e^{-i\beta}(321) - \frac23
P^C_{EW2,s}e^{-i\beta}(321) ~,
\eea
where the particle order for each diagram (top to bottom) is given in
parentheses. We have continued to label each diagram by an index
denoting the flavor of the popped quark pair, but under SU(3), these
are all equal. For $\bd \to K^+K^-K^0$, we take particle 1 as $K^+$,
particle 2 as $K^-$, particle 3 as $K^0$.  The amplitude is
\bea
A(\bd \to K^+K^0K^-) &=& -T'_{2,s}e^{i\gamma}(123) -
C'_{1,s}e^{i\gamma}(312) -{\hat P}'_{b;uc} e^{i\gamma}(123)
+ {\hat P}'_{b;tc}(123) \nn\\
&& \hskip-2.8truecm -~\frac13 P'_{EW1,u}(213) + \frac23 P'_{EW1,s}(123) - \frac13 P^{\prime C}_{EW1,u}(321) + \frac23
P^{\prime C}_{EW2,s}(321)  ~.
\eea
The penguin diagrams for the two decays are defined in
Eq.~(\ref{Pdefs1}). Comparing the two amplitudes, we see that, due to
$C_{1,s}$ and $P_{EW1,s}$, $\ket{A}$ and the mixed states are not
common. Therefore, the two decays are related only for $\ket{S}$.

Consider $\bs \to K^0 \kbar \kbar$ and $B^+ \to K^+ K^0 \kbar$. For
$\bs \to K^0 \kbar \kbar$, particle 1 is $K^0$, particles 2 and 3 are
$\kbar$ (consistent with the choice of mixed states above). $\ket{M_3}
= \ket{M_4} = \ket{A} = 0$.  The amplitude is
\bea
A(\bs \to \kbar \kbar K^0) &=& {\mathcal P}_{a;uc} e^{i\gamma} (213) +
{\mathcal P}_{a;tc} e^{-i\beta} (213) - \frac13 P_{EW1,s} e^{-i\beta} (123) \nn\\
&& \hskip-2.5truecm -~\frac13
P_{EW1,d} e^{-i\beta} (213) - \frac13 P_{EW1,s}^C e^{-i\beta} (213) - \frac13 P_{EW2,d}^C e^{-i\beta} (213)~.
\eea
For $B^+ \to K^+ K^0 \kbar$, we take particle 1 as $K^+$, particle 2
as $K^0$, and particle 3 as $\kbar$.  The amplitude is
\bea
\sqrt{2} A(B^+ \to K^+ K^0 \kbar) &=& {\mathcal P}'_{b;uc} e^{i\gamma} (231) - {\mathcal P}'_{b;tc} (231) + \frac13 P'_{EW1,s} (231) \nn\\
&& \hskip-2truecm +~\frac13 P'_{EW1,d} (321) + \frac13
P^{\prime C}_{EW2,s} (132) + \frac13 P^{\prime C}_{EW1,d}  (132) ~.
\eea
The penguin diagrams for the two decays are defined in
Eq.~(\ref{Pdefs2}).  Due to the EWP's, we see that the two decays are
related only for $\ket{S}$.

Consider $B^+ \to \pi^0 \kbar K^+$ and $B^+ \to \pi^0 K^0\pi^+$.  For
$B^+ \to \pi^0 \kbar K^+$, we take particle 1 as $\pi^0$, particle 2
as $\kbar$, and particle 3 is $K^+$. The amplitude is
\bea
\sqrt{2} A(B^+ \to K^+\kbar\pi^0) &=& -T_{1,s}e^{i\gamma} (321)- C_{2,s}e^{i\gamma} (321)
\nn\\
&& \hskip-3truecm +~{\mathcal P}_{b;uc}e^{i\gamma} (123) - {\hat P}_{a;uc}e^{i\gamma} (231) + {\mathcal P}_{b;tc}e^{-i\beta} (123) - {\hat P}_{a;tc}e^{-i\beta} (231) \nn\\
&& \hskip-3truecm -~P_{EW2,s}e^{-i\beta} (123) - \frac13 P^C_{EW1,d}e^{-i\beta} (321) - \frac23
P^C_{EW1,s} e^{-i\beta}(132) \nn\\
&& \hskip-3truecm +~\frac13 P^C_{EW2,u} e^{-i\beta}(132) - \frac13 P^C_{EW2,s} e^{-i\beta}(321) ~.
\label{BKKpiamp} 
\eea
The penguin diagrams are defined in Eqs.~(\ref{Pdefs1}) and
(\ref{Pdefs2}).  For $B^+ \to \pi^0K^0\pi^+$, we take particle 1 as
$\pi^0$, particle 2 as $K^0$, and particle 3 is $\pi^+$. The amplitude
is
\bea
\sqrt{2} A(B^+ \to K^0\pi^+\pi^0) &=& -T'_{1,d}e^{i\gamma} (321) -C'_{2,d}e^{i\gamma} (321) +
P'_{EW2,d}(123) \nn\\ 
&& +~\frac13 P^{\prime C}_{EW1,u} (312) + \frac23 P^{\prime
  C}_{EW1,d}(132) ~.
\eea
Under SU(3), ${\mathcal P}_{b} = {\hat P}_{a}$ Thus, in order for the
gluonic-penguin contribution to cancel in Eq.~(\ref{BKKpiamp}) above,
we need a state which is symmetric in $(123) \leftrightarrow
(231)$. This is $\ket{S}$ or $\ket{A}$ -- mixed states are
excluded. However, $\ket{A}$ is itself excluded by the $P^C_{EW1}$
contribution -- apart from CKM factors, it has the same sign in the
two amplitudes, despite the particle order being cyclic in one case
and anticyclic in the other. Thus, the two decay amplitudes are
related only for $\ket{S}$.

Finally, consider $B^+ \to \pi^- \pi^+ \pi^+$ and $B^+ \to \pi^- K^+
\pi^+$. For $B^+\to \pi^-\pi^+\pi^+$, particle 1 is $\pi^-$, particles
2 and 3 are $\pi^+$. This implies that $\ket{M_3} = \ket{M_4} =
\ket{A} = 0$.  The amplitude is
\bea
A(B^+\to \pi^-\pi^+\pi^+) &=& - T_{2,d} e^{i\gamma}(213) -~C_{1,d}e^{i\gamma} (231) - {\tilde P}_{b;uc} e^{i\gamma}(213) \nn\\
&& \hskip-3truecm -~{\tilde P}_{b;tc} (213) +~\frac13 P_{EW1,u} (123) -\frac23 P_{EW1,d} (213) \nn\\
&& \hskip-3truecm +~\frac13 P^C_{EW1,u} (213) -~\frac23 P^C_{EW2,d} (213) ~.
\eea
For $B^+ \to \pi^- K^+ \pi^+$, take particle 1 as $\pi^-$, particle 2
as $K^+$, and particle 3 is $\pi^+$. All six $S_3$ states allowed.
The amplitude is
\bea
A(B^+\to \pi^- K^+ \pi^+) &=& - T'_{2,d} e^{i\gamma}(213) -~C'_{1,d} e^{i\gamma} (231) - {\tilde P}'_{b;uc} e^{i\gamma} (213) \nn\\
&& \hskip-3truecm +~{\tilde P}'_{b;tc} (213) -~\frac13 P'_{EW1,u} (132) +\frac23 P'_{EW1,d} (312) \nn\\
&& \hskip-3truecm -~\frac13 P^{\prime,C}_{EW1,u} (312) +~\frac23 P^{\prime,C}_{EW2,d} (312) ~.
\eea
The penguin diagrams for the two decays are defined in
Eq.~(\ref{Pdefs1}). All states with $2\leftrightarrow 3$ symmetry are
allowed. Thus, unlike the above cases, the two decay amplitudes are
related for $\ket{S}$, $\ket{M_1}$ and $\ket{M_2}$. This is a special
case. Here, the processes are identical, save for the flavor of the
decay quark ($d$ or $s$). As a result, the amplitudes are equal for
all nonzero states. There is one other pair like this -- $B^+ \to K^-
\pi^+ K^+$ and $B^+ \to K^- K^+ K^+$. For all other pairs, the two
decay amplitudes are related only for $\ket{S}$ (or for all $S_3$
states in the case of U-spin pairs).

Now, in Refs.~\cite{diagramspaper,Kpipipaper} it was shown how the
$S_3$ states can be determined experimentally. Below we review the
method, focussing on the state $\ket{S}$.  Consider the decay $B^+ \to
\pi^+ K^- K^+$. The Dalitz-plot events can be described by $s_+ =
\left( p_{\pi^+} + p_{K^+} \right)^2$ and $s_- = \left( p_{\pi^+} +
p_{K^-} \right)^2$, so that the decay amplitude, ${\cal M}(s_+,s_-)$,
can be extracted. We introduce the third Mandelstam variable, $s_0 =
\left( p_{K^+} + p_{K^-} \right)^2$. It is related to $s_+$ and $s_-$
as follows:
\beq
s_+ + s_- + s_0 = m_B^2 + m_\pi^2 + 2m_K^2 ~.
\eeq
The totally symmetric SU(3) decay amplitude is then given by
\bea
\ket{S} &\!=\!& \frac{1}{\sqrt{6}} \left[ {\cal M}(s_+,s_-) + {\cal M}(s_-,s_+) +
  {\cal M}(s_+,s_0) \right. \nn\\
&& 
\hskip0.8truecm 
\left. +~{\cal M}(s_0,s_+) + {\cal M}(s_0,s_-) + {\cal
    M}(s_-,s_0) \right] ~.
\eea
The state $\ket{S}$ can be determined for the other decays
similarly. With this, the size of U-spin/SU(3) breaking can be
found through the measurement of $X$ using any of the SU(3)
pairs.

\subsection{Other signals of SU(3) breaking}

Finally, we note that there are certain decays which have identical
amplitudes for the totally symmetric state $\ket{S}$. They are given
by the processes within parentheses in the list in
Sec.~\ref{3bodynonUspin}.  For these, the branching ratios and direct
CP asymmetries should be equal in the SU(3) limit. Thus, by obtaining
the state $\ket{S}$ for these decays, the measurement of these
quantities constitutes a further test of SU(3) breaking.

\section{Conclusions}

Within U-spin symmetry ($d\leftrightarrow s$), the amplitudes for
certain charmless $\btod$ and $\btos$ decays are equal, apart from CKM
matrix elements. Using this, two methods were proposed for extracting
weak-phase information from measurements of particular U-spin decay
pairs. The theoretical uncertainty of these methods must include the
issue of U-spin breaking. In general, theoretical input is used to
address this. However, one of the points of the present paper is that
this breaking can be measured experimentally. Under U spin, the
branching ratios and direct CP asymmetries of the two decays are not
independent -- there is a relation among them. Thus, one can determine
U-spin breaking by measuring the four observables, and seeing the
extent to which this relation is not satisfied.

Furthermore, if one neglects annihilation- and exchange-type diagrams,
there are additional pairs of $B$ decays whose amplitudes are equal,
apart from CKM matrix elements. In this case, the symmetry is flavor
SU(3). Here, too, the relation among the four observables holds in the
SU(3) limit, so that SU(3)-breaking effects can be determined from
the measurements of these quantities.

In this paper, we present the list of two-body $B$ decay pairs from
which the size of the breaking can be obtained. In fact, there are
five such pairs for which these measurements have been done. We
present this data, along with the determination of U-spin/SU(3)
breaking. In many such decays, the calculation of the factorizable
contribution to the breaking is reliable. Taking this into account,
one can measure the size of nonfactorizable effects. It is expected
that these are small. However, there is one decay pair -- $\bd \to
\pi^+\pi^-$ and $\bd \to\pi^-K^+$ -- which exhibits large ($\sim
2.5\sigma$) nonfactorizable breaking. With only one data point, one
cannot draw any firm conclusions. However it does perhaps provide an
interesting hint, and raises questions about analyses which neglect
nonfactorizable U-spin/SU(3) breaking.

We also present the list of three-body $B$ decay pairs whose
amplitudes are the same, apart from CKM factors. However, here the
situation is more complicated. Under SU(3), the final-state particles
are all identical, and the equality of amplitudes holds (almost
always) only for the totally symmetric final state $|S\rangle$. Thus,
this state must be isolated experimentally in order to measure SU(3)
breaking, and we describe how to do this.

We discuss the decay pairs whose amplitudes are equal, including CKM
factors, within SU(3). For two-body decays, the size of SU(3) breaking
is indicated by comparing the branching ratios and direct CP
asymmetries of the two decays. For three-body decays, once again the
equality of amplitudes holds only for $|S\rangle$, so that this state
must be distinguised in order to probe SU(3) breaking.

Finally, we note in passing that the pure-penguin decay $\bs \to \kbar
\kbar K^0$ is particularly interesting. Here the final state is a CP
eigenstate. Thus, given that the direct CP asymmetry is expected to be
small, the measurement of the indirect CP asymmetry in this decay
cleanly probes the $\bs$-$\bsbar$ mixing phase. This might be easier
experimentally than performing the angular analysis in $\bs\to J/\psi
\phi$, which is what is done at present.

\bigskip
\noindent
{\bf Acknowledgments}: We thank A. Datta, M. Duraisamy, M. Gronau,
J. Rosner and D. Tonelli for helpful communications. This work was
financially supported by NSERC of Canada and FQRNT of Qu\'ebec.


\end{document}